\documentclass[fleqn,twoside]{article}
\usepackage{espcrc2}
\usepackage{epsfig}

\usepackage{graphicx}
\usepackage[figuresright]{rotating}

\newcommand{\EE}{\rm e^+ e^-}
\newcommand{\QQ}{\rm q \bar{q}}
\newcommand{\LL}{\ell^+ \ell^-}

\newcommand{\WW}{\rm W W^*}
\newcommand{\ZZ}{\rm Z Z^*}

\newcommand{\AmS}{{\protect\the\textfont2
  A\kern-.1667em\lower.5ex\hbox{M}\kern-.125emS}}

\title{Fermiophobic Higgs Bosons at LEP}

\author{A. Rosca\address[DESY]{DESY Zeuthen, 
        Platanenallee 6,  15738 Zeuthen, Germany}}
\begin{document}

\begin{abstract}
This work describes the results of the searches for a Higgs boson decaying 
into gauge bosons carried out by the four LEP collaborations: ALEPH, 
DELPHI, L3 and 
OPAL. A lower bound of 109.7 GeV is set at 95$\%$ confidence level on the
mass of a fermiophobic Higgs boson decaying into photons. This mass limit can 
be extended by considering the Higgs decay mode into weak bosons. Such a
combination has been done by the L3 collaboration which extended the 
fermiophobic mass by 5 GeV.
  
\vspace{1pc}
\end{abstract}

\maketitle

\section{Introduction}

Photonic final states from the process $\EE \to {\rm Zh} \to {\rm Z} \gamma 
\gamma$ do not occur in the Standard Model at the tree level, but may be 
present at a low rate due to charged weak boson and top quark loops.
Very specific models can lead to an enhancement of the h$\to \gamma \gamma$
rate, such as the fermiophobic 2 Higgs doublet models of type I [1], where all 
fermions are 
assumed to couple to the same scalar field, and the couplings can thus be
suppressed simultaneously by appropriate parameter choices.

Since there are several possible models which predict fermiophobic Higgs bosons,
the LEP Higgs working group (LHWG) defined the so-called "benchmark 
fermiophobic model" where
the production cross-section is the same as in the Standard Model, and the 
couplings of 
the Standard Model Higgs to fermions are removed, resulting in increased 
branching 
fractions into gauge bosons.
A fermiophobic Higgs in the benchmark model 
decays predominantly via the process h$\to \gamma \gamma$
if its mass is below 90 GeV, while at higher masses it decays mostly via the processes h$\to \WW, \ZZ$.

\section{Photonic Higgs decays}

A fermiophobic Higgs can be produced in $\EE$ collisions at the energies 
available at LEP by the Higgs-Strahlung mechanism in the s-channel, $\EE 
\to$ Zh, where the Higgs boson is radiated off an intermediate Z boson. 
For the case of photonic

\begin{table}[htbp]
\caption{Results of the h$\to \gamma \gamma$ search.}
\label{table:1}
\begin{tabular}{cccc}
\hline
Experiment      & Data & Bkgd. &  Mass
\\
           &  &  & 
limit \\
& & & (GeV) \\
\hline
ALEPH                & 23 & 30.8 & 105.4 \\
DELPHI               & 54 & 51.6 & 104.1 \\
L3                   & 62  & 72.0  & 105.4  \\
OPAL                 & 124 & 135.2 & 105.5 \\
\hline
SUM             & 263 & 289.6 & 109.7 \\
\hline
\end{tabular}
\end{table}

decays, this results in three categories of 
possible final states with different topologies, efficiencies and background 
rates:

\begin{figure}[htbp]
\vspace*{-10mm}
   \begin{center}
      \mbox{
          \epsfxsize=6.5cm
          \epsffile{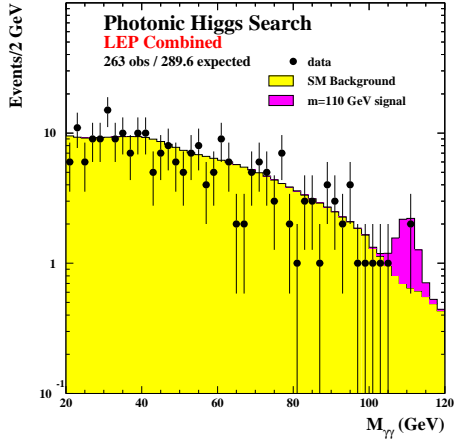}
           }
   \end{center}
\vspace*{-10mm}
\caption{Distribution of the reconstructed di-photon invariant mass for all
final states combined. The combined LEP data are shown together with the 
background and a Higgs boson signal with mass $m_{\rm h}$ = 110 GeV.
 }
\label{fig-1}
\end{figure}

\begin{enumerate}
\item 
the hadronic final state, involving two high energy photons and two jets from 
the hadronic decay of the Z;
\item
the missing energy final state, involving two high energy acoplanar photons 
and two neutrinos;
\item
the dilepton final state, involving two high energy photons and two high 
energy leptons.
\end{enumerate}

For all the channels, the major background is the double radiative Z 
production.  

The searches used data samples collected in $\EE$ collisions at centre-of-mass
energies between 88 and 209 GeV. The ALEPH [2], DELPHI [3], L3 [4] and OPAL 
[5] analyses are described in journal articles or CERN preprints.

The results of the analyses are summarised in Table 1, where we list
the number of observed events, the estimated background and the
lower mass limit set at 95$\%$ confidence level for the benchmark fermiophobic 
Higgs boson.

The distribution of the di-photon invariant mass is shown in Figure 1 for the combined data 
from the four LEP experiments. No excess with respect to the Standard Model predictions is
observed in the data. In the absence of a signal, the results are given in terms of upper limits
on the BR(h$\to \gamma \gamma)$ at 95$\%$ confidence level, assuming a Standard Model
production cross section for the Higgs boson. 
The LEP combined limits are shown in Figure 2.
In the fermiophobic model, the observed mass limit is 109.7 GeV.

\begin{figure}[htbp]
\vspace*{-10mm}
   \begin{center}
      \mbox{
          \epsfxsize=6.5cm
          \epsffile{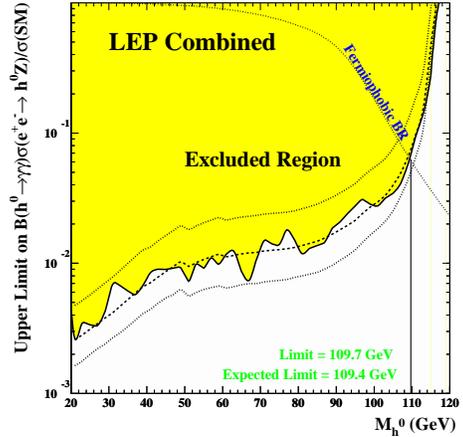}
           }
   \end{center}
\vspace*{-10mm}
\caption{Excluded values at 95$\%$ confidence level of BR(h$\to \gamma 
\gamma)\times \sigma(\EE \to {\rm Zh})/\sigma_{\rm SM}$ as a function of 
the
Higgs mass, in the fermiophobic model. The expected (dashed line) 95$\%$ 
confidence limit
and the theoretical prediction (dotted line) are also presented. 
   }
\label{fig-2}
\end{figure}

\section{Decay into weak bosons}

In the case Higgs decays via the process h$\to \WW$ or $\ZZ$, and 
considering
all the Z and W decay modes, there are nine possible channels. From these the
L3 [7] collaboration analysed the following six: Zh$\to \QQ \QQ \QQ$, 
Zh$\to \QQ 
\QQ \ell \nu$, Zh$\to \QQ \ell \nu \ell \nu$, Zh$\to \nu \nu \QQ \QQ$,
Zh$\to \nu \nu \QQ \ell \nu$ and Zh$\to \LL \QQ \QQ$, covering a total of 
92$\%$ of the h$\to \WW$ branching fraction.
Of these the qqqq$\ell \nu$ final state has the largest sensitivity.

The background in these channels comes from a variety of Standard Model 
processes, including W pair production, Z pair production and $\QQ$ events. 

The number of selected events in 336.4 pb$^{-1}$ of data collected at center-of-mass 
energies from 199 GeV up to 209 GeV is 566 and 568.3 events are expected from
the Standard Model processes. Since no indication for a signal was found in 
the data, the negative search result is given in terms of upper limits on the 
branching  fraction
BR(h$\to \WW$) as a function of the Higgs mass. The limits are 
shown in Figure 3.  A fermiophobic Higgs boson decaying to weak bosons is 
excluded for 83.8 GeV $<$ $m_{\rm h}$ $<$ 104.2 GeV with a region between
88.9 GeV $<$ $m_{\rm h}$ $<$ 89.4 GeV which can be excluded only at 93$\%$
confidence level.  

Model-independent fermiophobic results can be derived by scanning the relative
branching fractions of h$\to \gamma \gamma$ and h$\to \WW$, assuming
$$ 
{\rm BR}({\rm h} \to \gamma \gamma)+{\rm BR}({\rm h} \to \WW)+{\rm BR}({\rm h} 
\to \ZZ)=1. 
$$
The search results are presented as excluded regions in the plane [BR(h$\to 
\gamma \gamma$),$m_{\rm h}$] and are shown in Figure 4. The solid line 
crossing 
the excluded area gives the predicted value of BR(h$\to \gamma \gamma$) for 
the benchmark fermiophobic model. In the benchmark model, the 95$\%$ 
confidence level for a fermiophobic Higgs mass is set at 108.3 GeV, with an 
expected limit of 110.7 GeV. This result represents a significant 
extension of 
the mass limit of 105.3 GeV obtained by the L3 collaboration
from the search in the photonic decay mode.
The 95$\%$ confidence level mass limit, valid for any value of 
BR(h$\to \gamma \gamma$) and assuming a Standard Model production cross
section is set at 107 GeV.

\begin{figure}[htbp]
\vspace*{-5mm}   
\begin{center}
      \mbox{
          \epsfxsize=5.5cm
          \epsffile{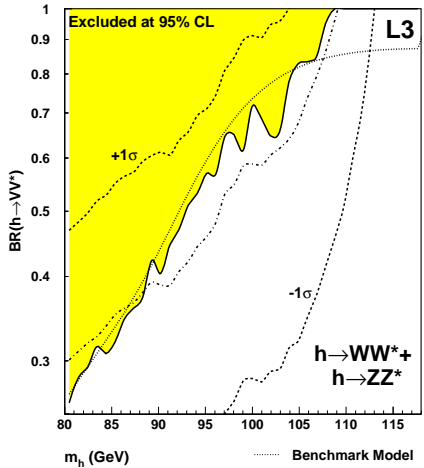}
           }
   \end{center}
\vspace*{-6mm}
\caption{Excluded values at 95$\%$ confidence level of BR(h$\to \WW)$ as a 
function of the Higgs mass in the fermiophobic model. The expected 95$\%$ 
confidence level limit (dot-dash line) and the theoretical prediction (dotted 
line) are also presented.
   }
\label{fig-3}
\end{figure}
\begin{figure}[htbp]
\vspace*{-5mm}
   \begin{center}
      \mbox{
          \epsfxsize=6.5cm
          \epsffile{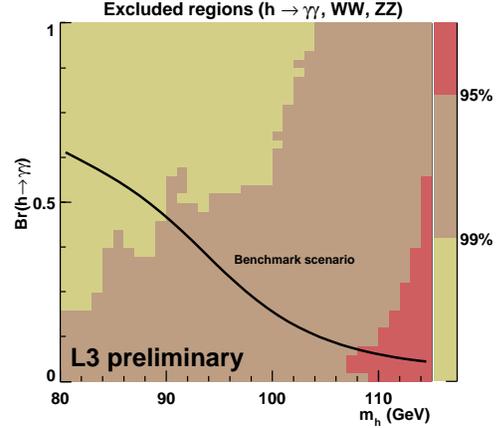}
           }
   \end{center}
\vspace*{-6mm}
\caption{Excluded area at 95$\%$ confidence level in the 
(BR(h$\to \gamma \gamma)$,$m_{\rm h}$) plane. The solid line gives the
prediction of the benchmark fermiophobic model.
   }
\label{fig-4}
\end{figure}

\section*{Acknowledgements}

I would like to thank the members of the four LEP collaborations and the LEP Higgs working group
for providing me with plots and their help in preparing this talk.

\end{document}